# In-Plane Bistable Nanowire For Memory Devices

B. Charlot[1,2], W. Sun[3], K. Yamashita[3], H. Fujita[3] and H. Toshiyoshi[2,3]
[1]IES CNRS/UMII, Place Eugène Bataillon, 34095 Montpellier Montpellier, FRANCE
[2]LIMMS CNRS/IIS, 4-6-1 komaba, Meguro-ku, 153-8505, Tokyo, JAPAN
[3]IIS, University of Tokyo, 4-6-1 komaba, Meguro-ku, 153-8505, Tokyo, JAPAN

*Abstract-* We present a micromechanical device designed to be used as a non-volatile mechanical memory. The structure is composed of a suspended slender nanowire (width: 100nm, thickness: 430nm, length: 8 to 30μm) clamped at both ends. Electrodes are placed on each side of the nanowire to 1) actuate the structure during the data writing and erasing mode and 2) determine its position by measuring the capacitive bridge in the reading mode. The structure is patterned by electron beam lithography on a pre-stressed thermally grown silicon dioxide layer. When later released by plasma etching, the stressed material relaxes and the beam buckles by itself to a position of lower energy. These symmetric bistable Euler beams exhibit two stable deformed. This paper presents the microfabrication process and the analysis of the static buckling of nanowires. Snapping of these nanowires from one stable position to another by mechanical or electrical means will also be discussed.

## I. INTRODUCTION

This paper presents a micromechanical device designed to be used as a static memory. The device is composed of a buckled nanowire that can be actuated between two stable positions in order to store information in a mechanical way. Random access memory (RAM) devices commonly used in computers are developed with microelectronics silicon technologies and can read, write, and erase data with low energy, low voltage and at high frequency. Various physical phenomena can be utilized to store data in memory devices. For dynamic RAMs and flash RAMs, electrical charges are held in a capacitor and floating-gate transistors, respectively. Flash RAMs are non-volatile memory, which allows information to be stored without a driving voltage.

Static RAMs operate in a different configuration. A storage cell consists of two inverters that are placed in a loop configuration forms an electrically bistable circuitry. A total of six essential MOS transistors occupies a relatively large device footprint, however it provides the fastest read/write time of about 0.4ns. The device presented in this paper employs a similar principle but uses mechanical bistable architecture rather than an electrical one.

Recently some new storage mechanisms have been developed to build advanced memory module. Magnetoresistive RAM, for example, store data as a resistive path. The structure is composed of a thin tunnel barrier sandwiched by two ferromagnetic layers where magnetization of one can be changed by applying an external field. Measuring the total electrical resistance of the layer stack provides the reading function. Ferroelectric RAMs are based on the use of a ferroelectric material (typically PZT) that can store a charge depending on the applied polarisation. The ferroelectric hysteresis effect, as seen in the charge versus voltage curve, enables data to be stored as a charge within the material.

With the advancements of micro and nano fabrication technologies, novel micro-mechanical structures designed as data storage medium have started to emerge. Research works on nanowires and Carbon Nanotubes (CNT) will be briefly discussed in the next section. The second section of this paper will describe these works. Section 3 explains in details the proposed bistable nanowire structure and the microfabrication process. In section 4, we attempt to analyze static buckling from a mathematical approach. We will also present some dimensional measurements of the self-deformed nanowire. In section 5, displacements of the nanowire due to mechanical probing or electrical actuation will be shown. Finally our findings will be concluded in section 6.

## II. MICROMECHANICAL MEMORIES

Previous micromechanical bistable designs [1][2][3] have quite large sizes. It is important to introduce new designs with cell footprint comparable to standard solid-state memory.

Some previous efforts such as [1][4][5][6] used stressed material as the buckling beam. These suspended silicon dioxide ribbon buckles out of plane upon sacrificial release.

Cavendish kinetics [7], has been developing micromechanical memory devices that use a clamped/free electrostatically actuated micro-cantilever as the storage medium. When pulled down in contact with a fixed electrode, the beam tip becomes stuck to electrode, which keeps the beam in a deformed position. By applying a voltage to another electrode, placed above the beam, it is possible to expel the beam and then to perform the erase function. It is recently reported that the process can be integrated on a CMOS 0.18μm technology to produce memory chips. Actuation voltage needed to write one bit can be as low as 1,8V, which is compliant with the CMOS technologies.

Recently another type of memory has been described in the literature [8][9][10]. This new structures uses Carbon Nano Tubes (CNT) as main part of the memory device. Several CNTs are clamped between metallic islands and are

 

suspended over a metallic electrode on top of an insulating layer. By applying a potential difference between the CNT and the metallic electrode, a generated electrostatic force can bend the CNT down towards the electrode, and this bent position can be maintained by some accumulated surface forces such as the Van der Waals force. Competition between the CNTs' restoring forces and the Van der Waals forces cause a bistable phenomenon. The reading function is achieved by measuring the current flowing trough the CNTs that are in contact with the metallic electrode. The erasing function is done by applying voltage to a second electrode located above the CNTs for pulling up. Although this system appears to be very promising, technical difficulties related to the deposition of CNTs at exact position keeps this technology at development stage.

### III. IN PLANE PRE-STRESSED NANOWIRE

We propose a structure based on the use of an in-plane bistable buckled nanowire. Figure 1 illustrates the schematic of the structure. It is composed of a clamped/clamped beam placed between two adjacent electrodes. The material is stressed during deposition, thus buckling occurs randomly in either direction upon sacrificial release. Data is stored as the mechanical position of the beam.

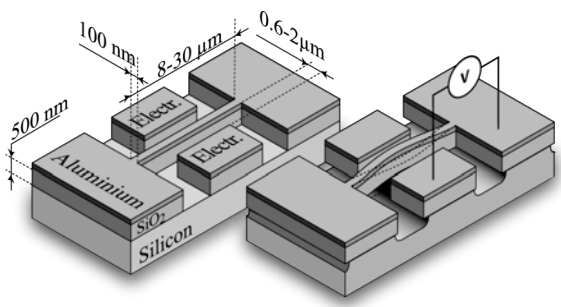

Fig. 1. Schematic of in-plane pre-stressed buckling beam to be used a non-volatile bistable mechanical memory.

By applying a sufficiently large driving voltage to a designated electrode, the electrically grounded nanowire will be electrostatically attracted and displaced to a stable position that is closer to the active electrode, thus achieving the writing or erasing function. The reading function is made by measuring the capacitance between the nanowire and one of the side electrodes. Thanks to the stability of the buckled structure, data can be stored permanently without the need of a sustaining voltage, therefore the device can be categorized as a non-volatile memory.

The main part of the micromechanical memory devices lies in the stressed layer, in which bistability is to be observed upon sacrificial release. Stress is induced to this layer by wet oxidation of silicon (100) wafer in an oxidation furnace at 1100°C as shown in Figure 2a. Ellipsometric measurements have shown that $SiO_2$ thicknesses range from 431 to 866 nm depending on the oxidizing duration. In order to buckle horizontally, the cross-section of the beam should be rectangular with a width smaller than the height. In order to obtain narrow beams of 100nm width, we have used the electron beam lithographic machine (E5112 from Advantest) available at the University of Tokyo VDEC facilities. The corresponding electron sensitive positive photoresist we used was ZEP520A-7, which is spin-coated at 5000 rpm, exposed under 120 $\mu C/cm^2$ dose, and then developed to reveal the pattern as illustrated in Figure 2b. After a rapid O2 ashing, a layer of 50nm aluminium is evaporated over the entire wafer surface as shown in Figure 2c. The resist is then removed by a lift off process to ensure only metallic patterns remain on the oxidized substrate.

With the metallic patterns over the $SiO_2$ layer as the etch mask, oxide etching is achieved using Reactive Ion etching (RIE) with a mixture of $CF_4/O_2$ gases (80/20sccm) for a pressure of 7 Pa and a RF power of 70W. A duration of 8 minutes is needed to etch anisotropically through the silicon dioxide layer, as shown in Figure 2d.

The next step is to etch the silicon structures isotropically in a $SF_6$ environment as shown in Figure 2e. We can take advantage of the underetching effect to dry-release the nanowires, which will buckle randomly in either left or right direction as a result. Additional steps of metal deposition are needed to cover sidewalls in order to increase the conductive surface for better electrostatic actuation of the beams. We tried several approaches including aluminium evaporation and tungsten sputtering. The latter provides conformal deposition to metalize also the sidewall, but a plasma environment within a sputtering chamber causes a charging effect especially at the edges and corners of a structure, which occasionally result in short-circuiting the beams and the side electrodes.

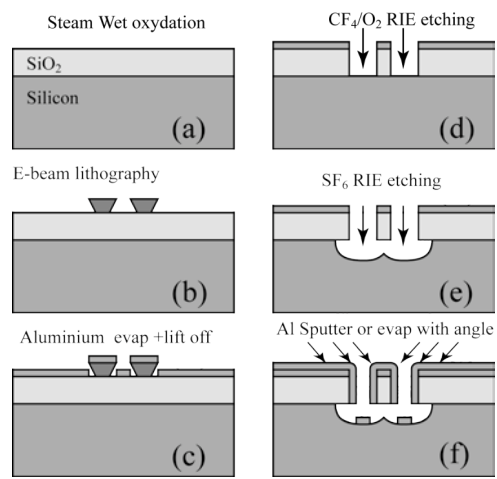

Fig. 2. Process flowchart

Figure 4 shows an SEM picture of the fabricated device. We can observe that for this particular device, the beam buckles to the left. The device shown has a length of 12 μm and a nominal width of 60 nm. Also, as can be seen in Figure 3b, we have a design that contains 4 electrodes, with two on each side of the nanowire, have been designed in order to study the influence of actuation position on the buckling behaviour of the beam.

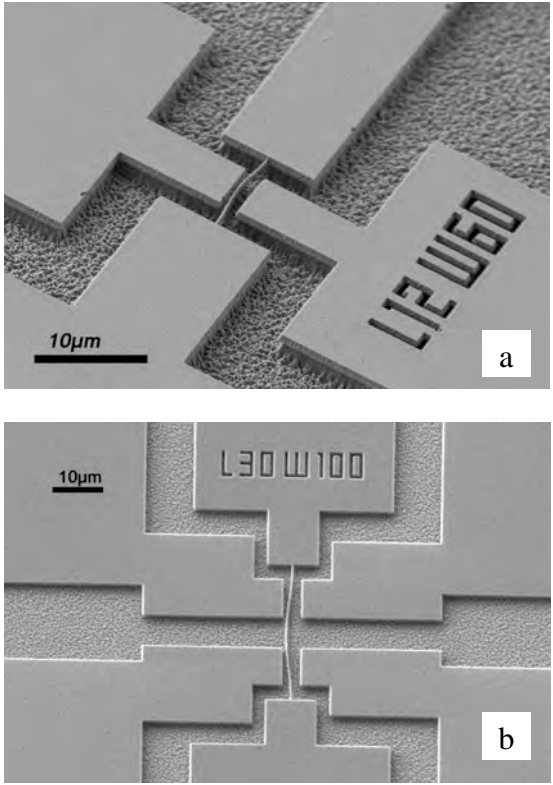

Fig. 3. SEM picture of the fabricated device showing 2 electrode (a) and 4 electrodes (b) structure.

## IV. STATIC BUCKLING

A classical formulation is the Euler-Bernouilli equation [11][12] where the equilibrium equation can be written as a fourth order spatial derivative of the lateral displacement $w(x)$:

$$\frac{d^4 w}{dx^4} + n^2 \frac{d^2 w}{dx^2} = 0 \quad \text{with} \quad n^2 = \frac{P}{EI} \quad (1)$$

where $P$ is the axial load, $E$ is the Young's modulus of the beam material and $I$ the moment of inertia of the beam. In the case of a clamped clamped beam, the following conditions apply:

$w(0) = w(l) = 0$ and $w'(0) = w'(l) = 0$

The solution of this equation is a set of orthogonal eigenfunctions $w_i(x)$ for which eigenvalues $n_i$ satisfies equation (2):

$$1 - \cos(n_i L) = \frac{1}{2} n_i L \sin(n_i L) \quad (2)$$

For even solutions (i=0,2,4,..) we have a cosine function:

$$w_i(x) = A_i \left[1 - \cos(n_i x)\right] \quad (3)$$

And for odd solutions (i=1,3,5,..) we have the following function

$$w_i(x) = A_i \left[1 - \cos(n_i x) - \frac{2}{n_i L}(n_i x - \sin(n_i x))\right] \quad (4)$$

Equations (3) and (4) are plotted in Figure 5 for i=0, 2, 4, and 1, 3, 5, respectively.

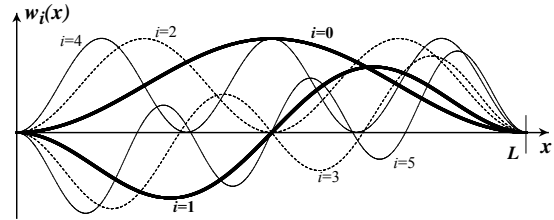

Fig. 4. Eigenfunctions of the Euler-Bernouilli buckling beam problem.

We can see in Figure 5, the buckled beam shape corresponds to first (*i*=0) and second (*i*=1) mode, respectively. The latter being observed in 30μm beam with 1μm inter-electrode distance. More derivations of this model were conducted by Timoshenko [13].

Concerning the stability of higher order modes, S.M Carr et al. [14][15][16] have shown that equivalent structures with mesoscale dimensions (length~10-20μm, width~200nm and thickness~500nm) can exhibit stable mechanical state composed of a superposition of several modes. Beams with high aspect ration (Length/width ration higher than 80) show stable superposition of modes. They also showed that first mode is dominant for an aspect ratio below 80.

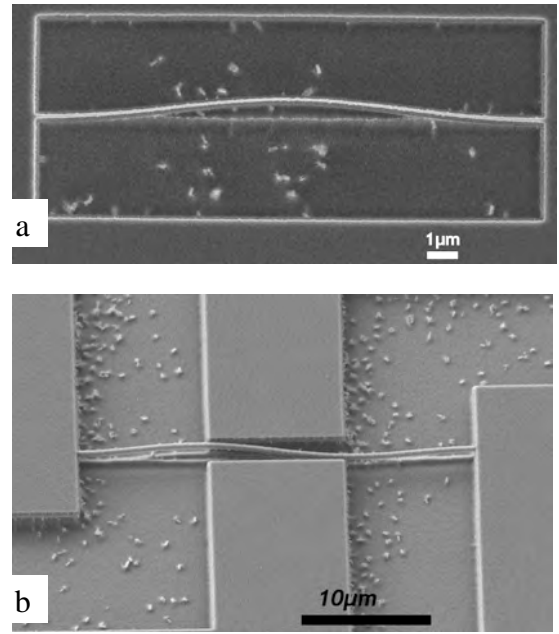

Fig. 5. FE-SEM image of first (a) and second (b) mode of the buckled beam.

In order to study the static buckling of beams, various designs with different beam dimensions have been included on a single wafer. For two nominal beam widths of 80nm and 100nm, we have included designs with beam lengths rang from 4μm to 20μm. The beam length L and the maximum deflection d at the centre of the buckled beams are measured by the Field Emission Scanning Electron Microscope (FE-SEM). The results for the two nominal widths are compared in Figure 6. The 4μm and 6μm short beams appear to have no buckle, and the deflections from the 8μm and 10μm beams are measured without high precision due to equipment limitations.

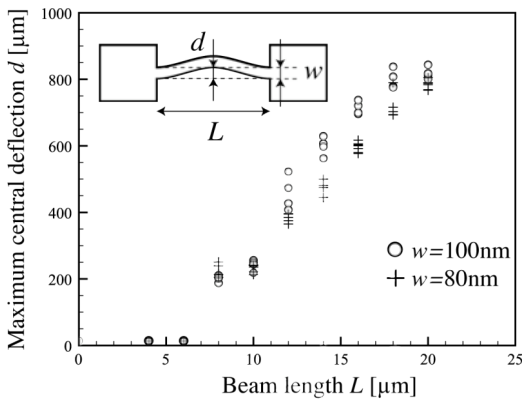

Fig. 6. Central deflection of bistable beams versus the beam length under two different nominal widths of 80 and 100 nm.

We also notice that the beam width is not uniform along the beam length, it has been measured to be about 60nm in the centre of the beam and about 120 nm near the edges. As shown in Figure 7, by using the secondary electron emission function of the FE-SEM, we can observe that the cross-section of the beam appears to be trapezoidal due to non-vertical sidewalls created by the plasma etching. The bottom width of the beam is measured to be about 260nm for 866nm thick beams.

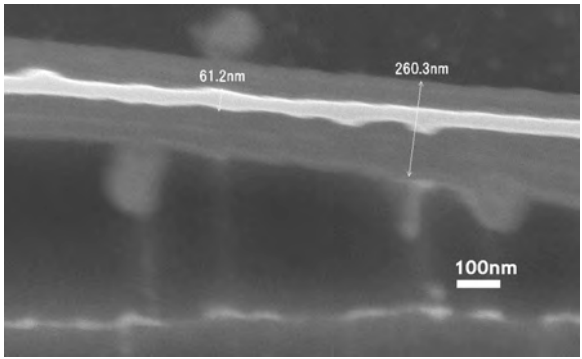

Fig. 7. SEM image of a buckled beam A close up view of a 20μm long, 100nm wide beam reveals a possibly trapezoidal cross-section.

In reference [1], Nicu et al. have developed a model for determining residual stress in $SiO_2$ layer by measuring the deflection of buckled beams. This method estimates the total potential and elastic energy of the beam based on the Euler-Bernoulli equation of the deformed beam. The internal compressive stress of the layer can be expressed in terms of the maximum deflection, the beam dimensions and Young's Modulus E. With this model, the critical stress, i.e the stress where a beam of given dimensions starts to buckle, is expressed in Equation (5):

$$\sigma_{cr} = \frac{\pi^2}{3} E \left( \frac{h}{L_0} \right)^2 \qquad (5)$$

Where $L_0$ and $h$ are the beam length and thickness, respectively. According to this model, a standard $SiO_2$ layer with about 270MPa compressive stress corresponds to a beam length of 3.4μm, where E and h are 69GPa and 120nm, respectively. This value can be compared with those reported in Figure 7, in which buckling can be observed for beams longer than 8μm.

In the work of Nicu et al. [1], the, $SiO_2$ beams have rectangular cross-sections and buckle out of plane. $SiO_2$ layer thickness was defined by oxidation time and could be precisely measured (by ellipsometry for example). In this work, the beams buckle laterally. The beam heights are known to be about 431nm and 866nm, but the beam width along the length as well as along the height are non-uniform due to various effects from the reactive ion etch. Moreover, the beam's moment of inertia I depends on the 3rd power of 1) width w for the lateral bucking case or 2) height h for the out-of-plane buckling case. Since these small variations in beam geometry causes large effects in the static buckling behaviour, discrepancy is expected if we model our beams with this mathematical expression.

V. MEASUREMENT

A. *Mechanical displacement*

In order to verify the bistable behaviour of the fabricated beams, we displaced the nanowire with a probe. It has been done using a test probe sliding on the surface of the wafer. The probe tip was first brought in contact with the substrate. Then it was drawn closer by sliding towards the nanowire under test. As shown in Figure 8a, the shadow on the right is actually the out-of-focus probe tip, which was then gently shifted horizontally towards the nanowire. As soon as contact was made, the nanowire snapped immediately to the second stable position. The displaced nanowire with the probe tip removed is shown in Figure 8b. Visible trails beneath the nanowire are remains of the anisotropic etching process.

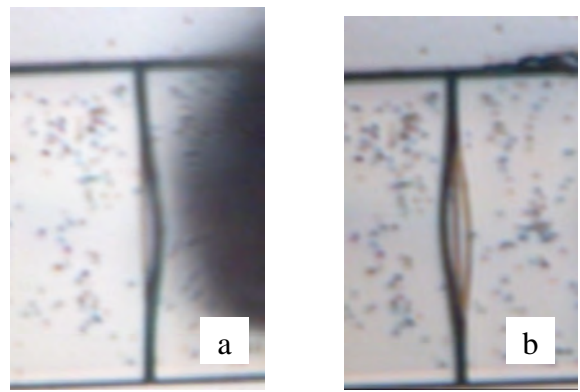

Fig. 8. Test probe displacement of a bistable beam. 100X optical microscope. (a) before probing and (b) after probing.

B. *Electrical actuation*

Electrical actuation of a 30μm-long beam with two separate electrodes on each side of the beam was conducted. While the nanowire is electrically grounded, a waveform generator and a high frequency amplifier are connected to provide a driving sinusoidal voltage at 90V peak to peak at one of the four available electrodes. The three remaining electrodes are kept electrically floating.

Figure 9 shows images of the buckling experiment captured by an overviewing CCD camera located on top of the microscope. Figure 9a shows the device before the experiment. The beam is naturally buckled towards the bottom of the page. By applying a driving signal to the upper left electrode and ramping up the peak-to-peak voltage, snapping of the beam can be observed at around 90V AC. The electrically actuated nanowire is shown in Figure 9b. Notice that the beam remains at the new stable position even after the voltage source is removed. Darkened corners around the target electrode indicate that an electrical breakdown might have occurred. In Figure 9c, the same driving voltage is applied to the lower left electrode, and the beam snaps back to the original position.

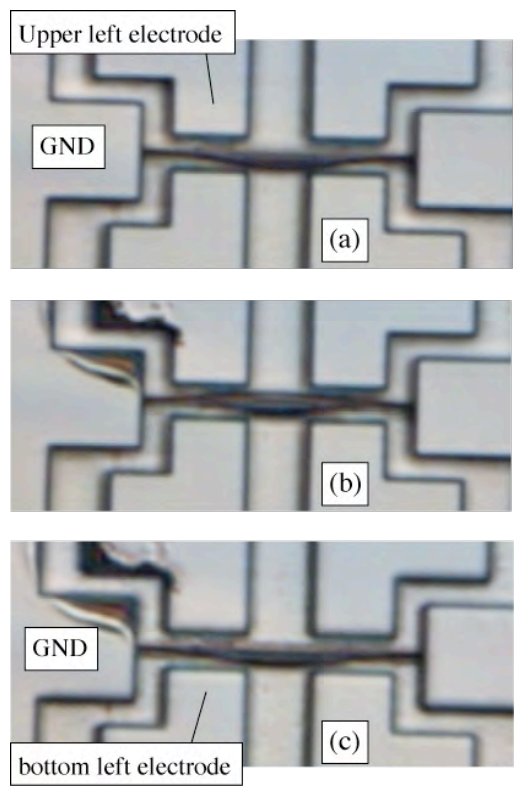

Fig. 9. Microscope images (100x) of a bistable beam under electrical actuation. (a) initial state, (b) actuation by upper left electrode, (c) restoration by bottom left electrode.

It is a major concern to lower the driving voltage in order to realize the device as memory modules. We are aware that the poor metallization along sidewalls provides inadequate fringe fields for effective electrostatic actuation. However, as mention previously, too well-conformed metallization would result in short-circuiting the entire structure. We are currently testing various metal deposition schemes and investigating if enlarging the isotropic envelop underneath the beams would help to widen the gaps to ensure better electrical isolation.

## VI. CONCLUSIONS

We have presented a micromechanical pre-stressed bistable beam structure to be used as a random access memory. The structure shows two stable positions that correspond to the first eigenmode. Switching from one position to another has been shown to be successful both by mechanical probing and electrical actuation. Due to insufficient sidewall coverage, high voltage of about 90V is required for electrical actuation. A continuation of this work will focus on the dynamic behaviour of the buckling mechanism, particularly the search of a frequency range within which the switching mechanism would be most effective.